\documentclass[12pt,floats,aps,showpacs]{revtex4}

\usepackage{amsmath}
\usepackage{amssymb}
\usepackage{stmaryrd}
\usepackage{epsfig}

\def\avg#1{\langle{#1}\rangle}

\newcommand{\BE}{\begin{equation}}
\newcommand{\EE}{\end{equation}}
\newcommand{\BA}{\begin{array}}
\newcommand{\EA}{\end{array}}
\newcommand{\beqn}{\begin{eqnarray}}
\newcommand{\eeqn}{\end{eqnarray}}

\newcommand{\etab}{\mbox{\boldmath $\eta$}}

\newcommand{\WAB}{W_{AB}}

\newcommand{\xf}{x_{\mathrm f}}
\newcommand{\vf}{v_{\mathrm f}}

\begin{document}
\baselineskip 26 pt
\title{
Discreteness effects in a reacting system of
particles with finite interaction radius.
}

\author{S.~Berti$^{1}$, C.~L\'opez$^{2}$, D.~Vergni$^{3}$ and A.~Vulpiani$^{4,5}$}

%\affiliation{}
\address{$^1$ Department of Mathematics and Statistics, University of Helsinki 
P.~O.~Box 68, FIN-00014 Helsinki, Finland.}
\address{$^2$ Unidad de F{\'\i}sica Interdisciplinar, IMEDEA 
(CSIC-UIB), Campus de la Universitat de les Illes
Balears, E-07122 Palma de Mallorca, Spain.} 
\address{$^3$ Istituto Applicazioni del Calcolo (IAC) - CNR,
Viale del Policlinico, 137, I-00161 Roma, Italy.}
\address{$^4$ Dipartimento di Fisica and INFN Sezione Universit\`a di Roma
``La Sapienza'', P.zle Aldo Moro 2, I-00185 Rome, Italy.}
\address{$^5$ INFM-SMC P.zle Aldo Moro 2, I-00185 Rome, Italy.} 

\date{\today}

%%%%%%%%%%%%%%%%%%%%%%%%%%%%%%%%%%%%%%%%%%%%%%%%%%%%%%%%%%%%%%%%%%%%%%%%%%%
\begin{abstract}

An autocatalytic reacting system with
particles interacting at a finite distance is studied.
We investigate the effects of the discrete-particle character
of the model on properties like reaction rate,
quenching phenomenon and front propagation,
focusing on differences with respect to the continuous case.
% typically described by
%the Fisher-Kolmogorov-Petrovskii-Piskunov (FKPP) equation.
We introduce a renormalized
reaction rate depending both on the interaction radius
 and  the particle density, and we relate it
to  macroscopic observables (e.g., front speed and
front thickness) of the system.

\end{abstract}
\pacs{05., 05.40.-a, 82.39.-k, 82.40.-g}
\maketitle
%%%%%%%%%%%%%%%%%%%%%%%%%%%%%%%%%%%%%%%%%%%%%%%%%%%%%%%%%%%%%%%%%%%%%%%%%%%%%%%%%%

\section{Introduction}
Most of the chemical and biological processes that appear in Nature
involve the dynamics of  particles (e.g., molecules or organisms)
that diffuse and interact with each other and/or with external
forces~\cite{murray,flierl,tel}.  If the total number of particles
per unit volume, $N$, is very large, a macroscopic description
of the system in terms of continuous fields, e.g., density or
concentration, is usually appropriate.
A prototypical model for these  
reaction-diffusion   
systems is the Fisher-Kolmogorov-Petrovskii-Piskunov (FKPP)
equation~\cite{kolmogorov,fisher} describing the spatio-temporal
evolution of a concentration
\begin{equation}
\partial_t \theta(x,t)=D \partial^2_{x x} \theta + p\theta(1 -\theta),
\label{fkpp}
\end{equation}
where $D$ is the diffusion coefficient, $p$ is the reaction rate that
determines the characteristic reaction time,
  $\tau=1/p$, 
and $\theta(x,t)$ is the concentration field (for simplicity 
we have assumed one spatial dimension).
It is well known~\cite{murray,saarloos,xin} that
Eq.~(\ref{fkpp}) admits uniformly translating solutions --fronts--
with a speed $v_0 = 2\sqrt{Dp}$ and a front thickness 
$\lambda_0 = 8\sqrt{D/p}$.
The above results do not depend
on the precise details of the reaction rule. Replacing
in Eq.~(\ref{fkpp}) $\theta(1 -\theta)$ with a convex function
$g(\theta)$, with $g(\theta)>0$ for $0<\theta<1$, $g(0)=g(1)=0$,
and $g'(0) = 1$, $g'(1)<0$, one has  
the same behaviour for $v_0$ and $\lambda_0$~\cite{aronson}.

On the other hand, if the number of particles per unit volume
is not very large, the continuous description could
not be appropriate. In such a case, one can consider
a discrete particle model with  
$N$ particles whose positions ${\bf x}_\alpha(t)$
evolve 
according to the Brownian motion 
\begin{equation}
\frac{d{\bf x}_\alpha(t)}{dt}=\sqrt{2D}\etab_{\alpha}\,\,,
\,\,\, \alpha=1,...,N,
\label{eq:bmcontinuous}
\end{equation}
where $\etab$ is a white noise term.
Moreover each particle is characterized by a  {\it color} $C_{\alpha}(t)$
which determines the particle type.
The model is completed by  the
reaction {\it rule} between particles. 
In order to obtain an autocatalytic reaction
\begin{equation}
 A + B \,\, \stackrel{ p }{\longrightarrow} \,\, 2B, 
   \label{eq:autocatalytic}
\end{equation}
one can consider just two types of particles 
$C=0$ (unstable) and $C=1$ (stable), that correspond to the species
$A$ and $B$, respectively, with the following 
dynamics: particles  of type $1$ always remain 
$1$, and particle $0$ changes to $1$ with 
a given probability that depends
both on $p$, the reaction rate,
and on how many $1$ particles are around it.
It is not difficult to realize that in a suitable continuum
limit, Eq.~(\ref{fkpp}) gives the evolution of the
{\it color concentration} of this microscopic system
(see Section~\ref{sec:model}).
The aim of this work is precisely to study the
case in which the density of individuals is  small,
and therefore the discrete nature of the system can play a role
\cite{young,lopez}.

Several approaches have been adopted 
to investigate the relevance of the correction
to the continuum limit.
On one side,
it has  been
assumed that the 
dynamics of the system is given by deterministic macroscopic
equations like Eq.~(\ref{fkpp}), and
a noise term, of order $1/\sqrt{N}$,  
which  takes into account the
microscopic fluctuations originated by
the finite number of particles~\cite{discrete1}.
On the other side, 
following the work of
Brunet and Derrida~\cite{brunet}, this problem
has been successfully studied by  using a cutoff
at the density value $1/N$ for the continuous
field equations.
This has been employed to determine
corrections to some front properties in FKPP-like
equations (see \cite{panja} for a review). 
In particular, it has been shown
that the deviation from the continuum
value of the front speed is of the
order $1/(\ln N)^2$, which is
rather significant~\cite{brunet}.

More recently,
Kaneko and coworkers~\cite{kaneko}
analyzed the dynamics of some chemical
reactions, studying the influence of the 
molecular discreteness.
They identify typical length scales in the system
which may separate the {\it continuum} behavior
from the {\it discreteness-influenced} one.
They report transitions 
to a novel state with symmetry breaking that
is induced by discreteness, but they do
not investigate the features of
front propagation  properties
in terms of the number of particles.
A crucial quantity  
 is the so-called Kuramoto length, 
$l_K =\sqrt{2 D \tau}$, which is proportional to
the front thickness and measures the typical distance 
over which an unstable particle diffuses during its lifetime
(note that $\tau=1/p$ can be interpreted as the  average 
lifetime, i.e., the time particles live before they react).
In some situations, especially when there is a propagating front,
if the typical distance among particles is much smaller 
than $l_K$, the concentration of the particles 
can be regarded as continuous. On the other hand, when there 
are not many particles within a region of size $l_K$, 
discreteness effects should be taken into account~\cite{kaneko}.

In our work we study the interplay between length scales in the
problem, our principal aim being to explain the effects of the
discrete nature of the system on properties like reaction rate,
quenching and front speed.  
Differently from most of the works
in  discrete reaction-diffusion systems,
we do not consider a lattice model: particles diffusively move in space 
and interact when their distance is smaller than an interaction radius $R$,
which corresponds to a 
natural length-scale
appearing in many chemical and biological systems~\cite{plankton,lopez}. 
We study several properties of the system as
a function of $R$, realising, via comparison of different length-scales,
when the effects of discreteness have a dominant role. 
As expected, the continuum limit is described by the
FKPP equation. Nevertheless we remark that in order to have the proper 
continuum limit it is not sufficient to have a very large density of particles.
We discuss the problem in the framework of chemical reaction dynamics, 
but everything can be thought in the
context of  population dynamics.

The Paper is organized as follows. In the next section we
present the particle model for the autocatalytic reaction. In section
\ref{sec:premixed} we study the renormalized reaction
rate of the system when particles of both types are in a closed
vessel, initially uniformly random distributed in space.
In section \ref{sec:quenching}  we study 
quenching phenomena when $B$ particles can turn into $A$ particles;
this causes the emergence of new properties of the model
that will be studied in detail.
Then,
in Sect. \ref{sec:front} we investigate the
front properties of the model (by choosing a proper initial
distribution and considering an infinite system in the
propagation direction), mainly in terms of
the interaction radius of the system.
Section \ref{sec:conclusion} presents
our conclusions.

\section{Model}
\label{sec:model}
Consider $N$ particles in a two-dimensional box
of size $L_x \times L_y$. Each particle
is identified by its position, ${\bf x}_\alpha(t)$, and its {\it color},
$C_\alpha(t)$, indicating the particle type.
To specify the dynamics it is necessary to give the evolution
rule for the position and the interaction rule between particles (chemistry).
Space will be considered continuous while time will be discrete (with
time step $\Delta t$). 
However its value, if small enough, it is not relevant.
Particle dynamics is synchronous, i.e., all
particle properties are updated at the same time.

The position evolution is given by 
\begin{equation}
	{\bf x}_\alpha (t+\Delta t) = 
	{\bf x}_\alpha (t) + \sqrt{2D{\Delta} t}\, {\bf u}_\alpha(t)\,\,,
\qquad \alpha=1,...,N,
	\label{eq:bmdiscrete}
\end{equation}
where $D$ is the diffusion coefficient, 
${\bf u}_\alpha(t)=(u_{\alpha,1} (t),u_{\alpha,2}(t))$
are stochastic Gaussian variables with the properties
$\avg{{\bf u}_\alpha(t)} = 0$ and 
$\avg{u_{\alpha, i}(n\Delta t)u_{\alpha, j} (m\Delta t)} =
\delta_{ij} \delta_{\alpha\beta} \delta_{mn}$, i.e.,
particles perform a discrete-time Brownian motion.

As already mentioned, 
to model an autocatalytic reaction (\ref{eq:autocatalytic}),
we consider two kinds of particles:
type $A$ particles, $C_\alpha=0$ (unstable), 
and type $B$ particles, $C_\alpha=1$ (stable).
The chemical evolution of the particles is 
given by the following stochastic process:
\begin{itemize}
\item if $C_\alpha(t) = 0$ then $C_\alpha(t+\Delta t) = 1$ with
  probability $P_{AB} = W_{AB}  \Delta t$;
\item if $C_\alpha(t) = 1$ then $C_\alpha(t+\Delta t) = 1$.
\label{colordynamics}
\end{itemize}
The probability (per unit time) $W_{AB}$
depends on the number of stable particles within the interaction 
radius. In fact, in the continuum limit, the autocatalytic 
reaction (\ref{eq:autocatalytic}) is expected to obey the mass action law
${\displaystyle {{\mathrm d} \Theta_A \over {\mathrm d}t}} = 
-p  \Theta_A  \Theta_B$, where $\Theta_A$ and $\Theta_B$
are the concentrations of particles $A$ and $B$, respectively, 
with $\Theta_A +\Theta_B =1$.
The probability that a  particle $A$ changes into a $B$ particle
is assumed to be 
\begin{equation}
   \WAB = p  \frac{N_R(B)}{\langle N_{loc} (R) \rangle}
        = p \frac{N_R(B)}{\pi R^2 \rho} \,\,.
   \label{eq:probreaction}
\end{equation}
where $N_R(B)$ indicates the number of $B$ 
particles within the interaction radius $R$ around the given particle $A$, 
$\langle N_{loc} (R) \rangle$ is the spatial
average number
of particles (of any type) in a radius $R$, and
$\rho = N/(L_xL_y)$ 
is the  density of particles. 

We discuss in the following that in 
a suitable limit the previous probabilistic rule 
converges to the FKPP equation. Let 
$N(A,t)$ and $N(B,t)$ be the total number of $A$ and $B$ particles, 
respectively; of course $N=N(A,t) +N(B,t)$ is constant.
The dynamics of the number of B particles
is given by the discrete stochastic process
\begin{equation}
N(B,t + \Delta t) = N(B,t) + \sum_{k \in N(A,t)} y_k\,,
\label{eq:difference}
\end{equation}
where $k$ is the index identifying $A$ particles and
$y_k$ is a discrete random variable which is 1
with probability $\Delta t \WAB$ (when the particle $A$ changes
into a $B$ particle), and is $0$  with probability $1-\Delta t \WAB$ 
(when the particle $A$ remains $A$).
For the expected value of $N(B,t)$, one has
$$
E(N(B,t + \Delta t)) = E(N(B,t)) + E(N(A,t))  p  
                      \frac{E(N_R(B,t))}{\pi R^2 \rho} \Delta t  = $$
$$\qquad =            E(N(B,t)) + p [N - E(N(B,t))]  
                      \frac{E(N_R(B,t))}{\pi R^2 \rho} \Delta t\,.
$$
After a little algebra we obtain 
\begin{equation}
\frac{{\mathrm d}}{{\mathrm d}t}\Theta_B(t) = \lim_{\Delta t \to 0}  
  \frac{\Theta_B(t + \Delta t) - \Theta_B(t)}{\Delta t} = 
  p  (1 - \Theta_B(t)) \frac{E(N_R(B,t))}{\pi R^2 \rho} \,.
\label{eq:DiscreteFKPP}
\end{equation}
where $\Theta_B = E(N(B,t)) / N$ indicates the expected average 
concentration of B particles. In the case of an infinite number
of spatially premixed  particles the last term on the right-hand-side
of the above
relation becomes $\Theta_B(t)$ and we finally obtain the FKPP
equation for the homogeneous case: 
\begin{equation}
\frac{{\mathrm d}}{{\mathrm d}t}\Theta_B(t) = 
p (1 - \Theta_B(t)) \Theta_B(t)\,.
\label{eq:SingleParticleFKPP}
\end{equation}
In general,
under non-premixed spatial
conditions and/or a small density,
${\displaystyle {\frac{E(N_R(B,t))}{\pi R^2 \rho}}} \neq \Theta_B$ and 
the system  cannot be described by the  
FKPP dynamics.

Concerning the relevant length scales of the system one can 
identify the following ones: 
i) the mean nearest neighbour distance between particles, 
   $d_m = {\displaystyle {1 \over 2\sqrt{\rho}}} =
   \sqrt{{\displaystyle {L_x L_y \over 4 N}}}$ , 
ii) the interaction radius of the model, $R$, 
iii) the Kuramoto length scale, $l_K$ and
iv) the size of the system $L$. 
It is expected that the continuum limit is obtained when 
$d_m \ll R \ll l_K \ll L$. While the scale separation between
$d_m$ and $L$ can be easily achieved, in many situations
it might happen that the condition $R \ll l_K$ is not verified, 
or that $R$ is of the same order of $d_m$. In this
case the evolution of the system could be very
different from that of the continuous FKPP limit. It is
the objective of this work to investigate some properties of
the model in this regime.

Before starting with the discussion of the numerical
results, some comments
follow about the role of diffusion.
Since we introduce
the natural length-scale of the interaction, $R$, 
a diffusive time related to this distance arises
$t_D(R) = R^2 / D$.
When this time is much smaller than the reaction time $\tau=1/p$
the system is locally homogeneized before reaction happens.
In order to focus on the reaction properties rather than on the
diffusive effects we work in the limit
$t_D \gg 1/p$.

\section{Premixed particles in closed basins}
\label{sec:premixed}

Firstly we study the model in a closed vessel, where,
as initial condition, particles of both types are premixed 
and uniformly randomly distributed in space. 
In such a  case, 
the system evolution 
necessarily ends with the complete filling of the  box 
with type $B$ particles. Therefore the most significant 
physical quantity is the filling rate of particles $B$,
which is related to the reaction rate.
We proceed by fixing the value of $R$ and varying $N$
in order to explore different situations: 
a) continuum limit, $d_m \ll R$; b) the effect of the
discreteness, $d_m \gtrsim R$. 
In this case, at variance with front propagation properties discussed
in Sect.~\ref{sec:front}, we will see that the Kuramoto length  
does not play a fundamental role.
 The basic reason for this is
the spatially random distribution of particles. 

We adopt periodic boundary conditions on a square domain of side
$L_x=L_y=1$; the reaction rate is set to $p=1$; and averages are
numerically computed over a large number of noise realizations.

\begin{figure}[h]
\epsfig{figure=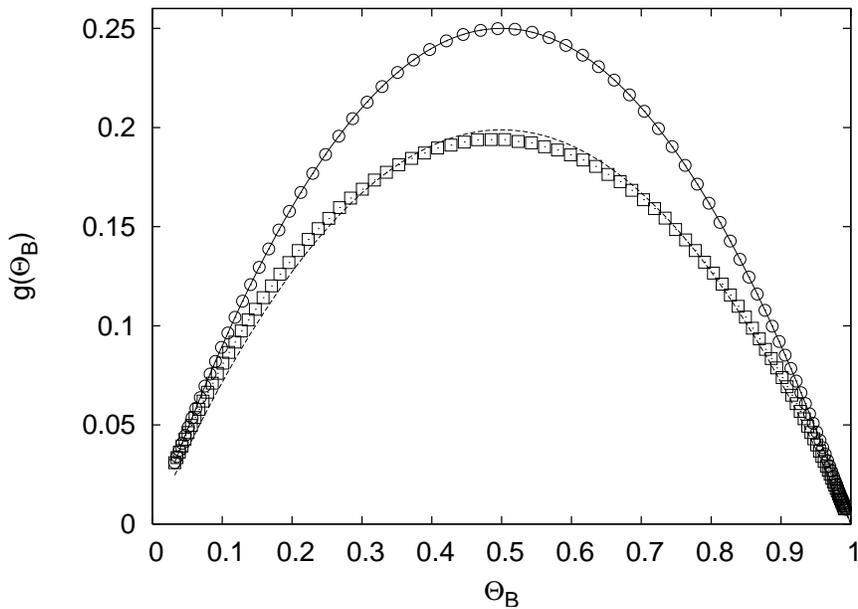,clip=true,width=12cm,angle=0}
\caption{The growth rate $g(\Theta_B)$ vs
 $\Theta_B$ (see Eq.~\ref{eq:DiscreteHomogeneousFKPP})
for $N=1000$
($\boxempty$),  and $N = 100000$ ($\circ$)
with $p = 1$,
$D=0.001$, $R=0.05$; the initial concentration corresponds to
$97\%$ of type $A$ particles and $3\%$ of $B$ particles,
uniformly distributed.
The solid and the dashed lines correspond to 
the quantity $p_R (N) \Theta_B (1-\Theta_B)$,
where $p_R (N)$ is fitted from numerical results
of the discrete particle model:
solid-line is for $p_R(N)=1$ and
dashed-line for $p_R(N)=0.8$.}
\label{fig:fitg}
\end{figure}

Since particles are well premixed, the process is spatially
homogenous, and we may assume that the growth rate of $\Theta_B(t)$, 
(see Eq.~\ref{eq:DiscreteFKPP}) is 
\begin{equation}
g(\Theta_B)=
p  (1 - \Theta_B(t)) \frac{E(N_R(B,t))}{\pi R^2 \rho}
\label{eq:DiscreteHomogeneousFKPP}
\end{equation}

In  the case of large particle density 
one expects that $g(\Theta_B) = p \Theta_B(1-\Theta_B)$,
therefore it is natural to assume  that for finite $N$
one can replace eq.~(\ref{eq:DiscreteHomogeneousFKPP})
with 
\begin{equation}
g(\Theta_B) =
p_R(N) \Theta_B(1-\Theta_B),
\label{renormalized}
\end{equation}
 where $p_R(N)$ is a renormalized reaction
rate of the  discrete particle model.
In this way the evolution of $\Theta_B$ is given by an
FKPP equation with a {\it renormalized} ($R-$ and $N-$~dependent) 
reaction probability,
where  $\tau_R(N)=1/p_R(N)$ is the renormalized reaction
time for the system.
 Note that 
$p_R(N)$ contains all of the dependence of our system on 
the interaction radius and the number of particles and, therefore, it
is the proper quantity to study the influence of the discreteness
in the model.

 In Figure~\ref{fig:fitg} it is shown, for a given $R$,
the function $g(\Theta_B)$, with the approximation
of Eq.~(\ref{renormalized}) for two different values of $N$.
With the appropriate $p_R(N)$ value, the fit is rather good
and, for large $N$,  $p_R(N) \to p$. 

\begin{figure}[h]
\epsfig{figure=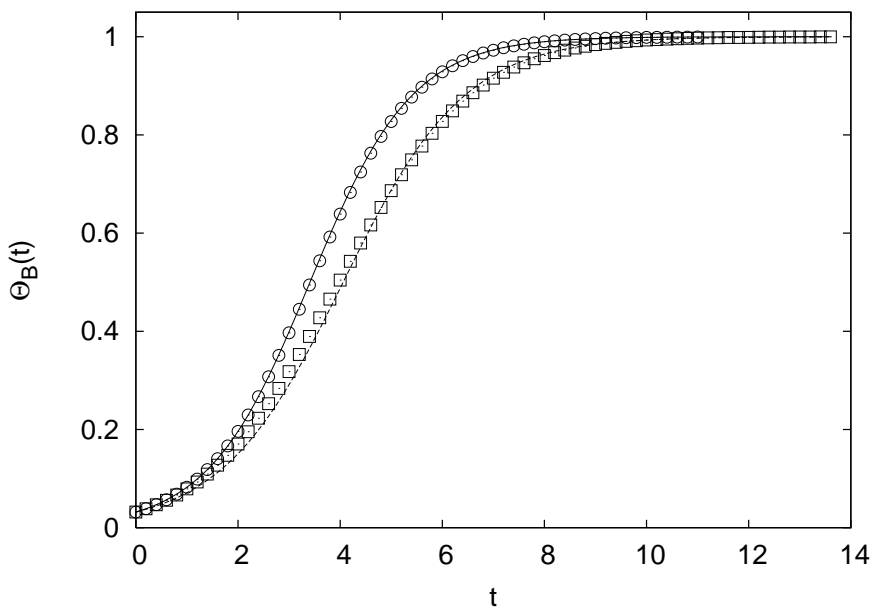,
clip=true,width=12cm,angle=0}
\caption{$\Theta_B (t)$ versus $t$ in the same experimental asset
(and using the same symbol) of Fig.~\ref{fig:fitg}. 
The solid and the dashed lines correspond to 
the fit of Eq.~(\ref{sol1}) on the experimental measures.
In particular solid-line is for $p_R(N)=1.0$ and
dashed-line for $p_R(N)=0.82$.}
\label{fig:fit}
\end{figure}

The equation ${\displaystyle {{\mathrm d} \Theta_B \over {\mathrm d}t}}
 = p_R (N) \Theta_B (1 -\Theta_B)$
can be easily solved: 
\begin{equation}
\Theta_B (t) = \frac{\Theta_B (0) e^{p_R(N) t}}{1 + \Theta_B (0) (e^{p_R(N) t} -1 )}.
\label{sol1}
\end{equation}

\begin{figure}[h]
\epsfig{figure=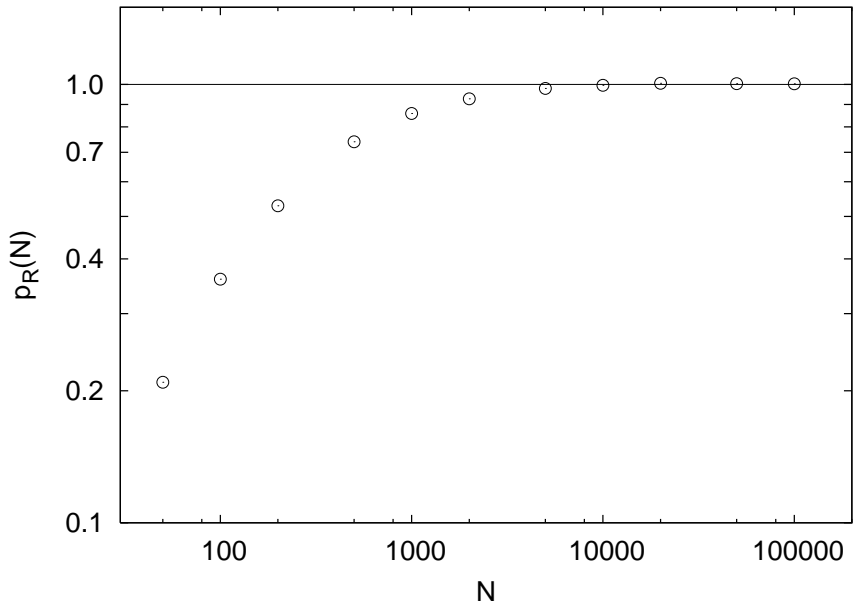,width=12truecm,angle=0}
\caption{ $p_R(N)$ versus $N$. The parameters are
the same as in Fig. 1. In particular, $R=0.05$, and 
the continuum limit is obtained for $N \approx 1000$ for which 
$d_m \approx 0.015$.}
\label{fig:probrenorm}
\end{figure}

Thus looking at the evolution of $\Theta_B =E(N (B,t))/N$ and using
Eq.~(\ref{sol1}) we have a value of $p_R(N)$ which is, in principle,
different from the one in Eq.(\ref{renormalized}).
However, the two values are rather close, and in the
following we will present results only for the
last one, obtained from Eq.(\ref{sol1}).
As an example, in Fig.~\ref{fig:fit}, we show 
 $\Theta_B(t)$ versus time obtained from the numerical
simulation of the particle model, and the best fit using
Eq.~(\ref{sol1}) from which a value of $p_R(N)$ comes out.
As previously shown in Figure~\ref{fig:fitg}, 
for large $N$ the value of $p_R(N)$ goes to the
continuum limit $p$. In  Figure~\ref{fig:probrenorm}, 
where the renormalized reaction probability versus $N$ 
is plotted, one sees that 
the continuum limit, $p_R(N) = p $ is, as expected,
obtained with good accuracy for large $N$ values.
This corresponds to 
  $d_m \ll R$,
while for values of $N$ such that  $d_m$ is comparable or smaller
than $R$ the continuous description becomes inaccurate. 

\begin{center}
\begin{figure}
\epsfig{figure=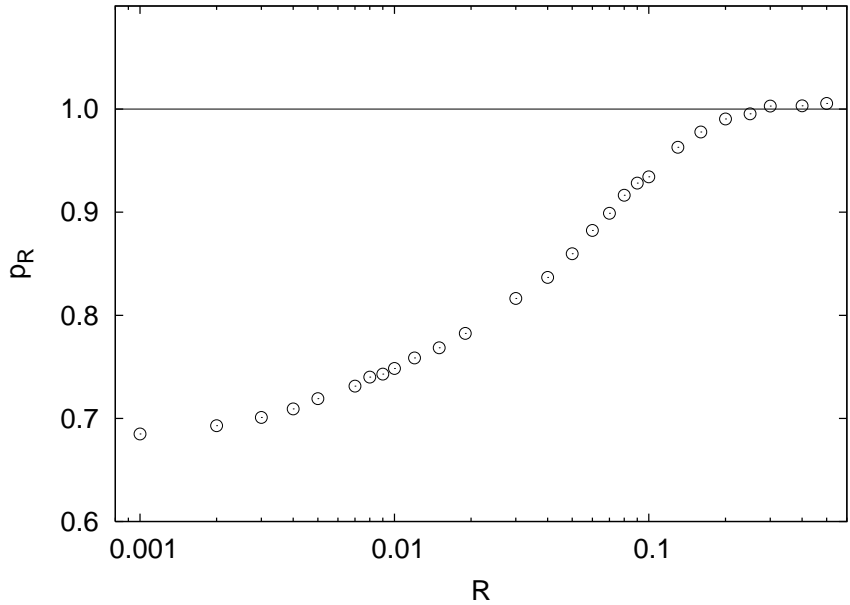,clip=true,width=12truecm,angle=0}
\caption{The renormalized reaction probability $p_R(N)$ versus $R$,
 using the fit of equation~(\ref{sol1}). 
$N=1000$, $l_K =0.045$ and  $d_m=0.0158$.}
\label{fig:versusR}
\end{figure}
\end{center}

More important for our purpose is the behavior of $p_R(N)$ versus
$R$. With a fixed total number of particles, $N$,
and a  well premixed initial condition,
we compute $p_R(N)$ varying the interaction radius $R$ (see
fig.  \ref{fig:versusR}).  We observe that in the continuum limit
($d_m \ll R$) we recover $p_R(N)=p$.  For small $R$, such that $d_m >
R$, $p_R(N)$ seems to reach a constant value, which is around $30 \%$
smaller than the FKPP one.

Few words have to be spent about the difference between
the large  $N$  limit (of Figure~\ref{fig:probrenorm})
and the large $R$  limit of Figure~\ref{fig:versusR}.
For the problem under discussion one has the same
behavior of the continuum limit if $d_m \ll R$
irrespective of the value of the Kuramoto length.
For example, in Fig.~\ref{fig:versusR}
one has $l_K \approx 0.045$
which is much smaller than the values of $R$ for which
the {\it continuum} limit holds. 
On the other hand, in the 
study of front properties we will see that 
the scenario is different and $l_K$ can play
a relevant role.

\begin{center}
\begin{figure}
\epsfig{figure=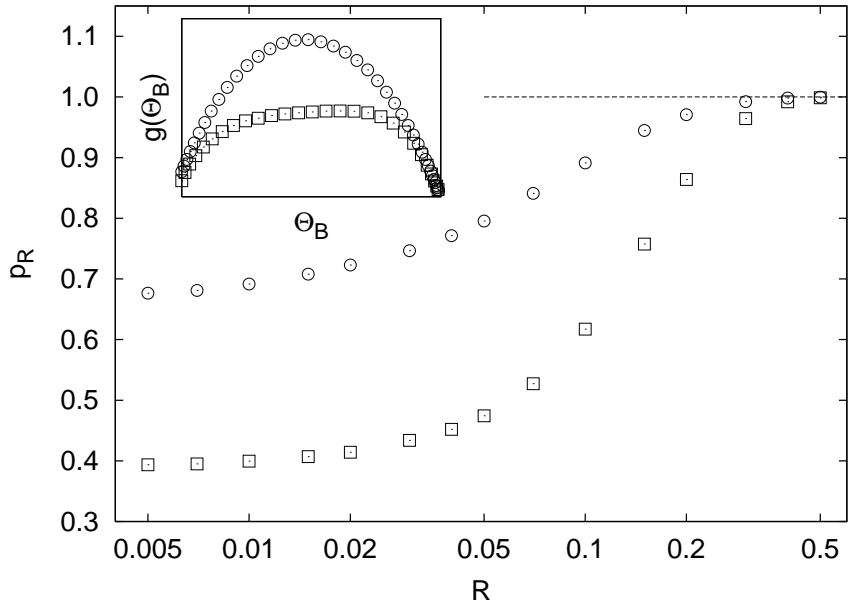,clip=true,width=12truecm}
\caption{$p_R(N)$ versus $R$ for different initial conditions:
$(\circ)$ premixed, $(\boxempty)$ B particles on the left of the
system and A particles on the right. The parameters are the same as 
in fig. \ref{fig:versusR}.  In the inset we plot the corresponding
$g(\Theta_B)$ vs $\Theta_B$ for $R=0.05$; $(\circ)$ premixed, and
$(\boxempty)$ initially separated particles.}
\label{fig:compinitcond}
\end{figure}
\end{center}

Now we want to discuss the 
dependence of the previous results on the chosen initial condition.
In Figure~\ref{fig:compinitcond}
we compare $p_R(N)$ in the premixed case, and when particles
are initially  separated in space.
Indeed, if the premixed particle condition is relaxed,
and the system is initially 
prepared with different spatial distributions
of particles, the renormalized reaction probability significantly 
changes, since $E(N_R(B,t))$ strongly depends 
on the particle configuration (see the inset
in Figure~\ref{fig:compinitcond}). 

\section{Possibilities of Quenching.}
\label{sec:quenching}

Studies on the quenching phenomenon~\cite{Constantin}
shows that in a continuous reaction-diffusion system 
in  presence
of an advectin velocity field,
and with a reaction term  of ignition type, i.e.
$g(\Theta)=0$ for $\Theta <\Theta_c$,
for a suitable size 
 of the ``hot'' region,
there is no propagating front, and the reaction quenches.
On the contrary,
 in a continuous FKPP 
system~(\ref{fkpp}) quenching phenomena do not appear~\cite{Roquejoffre}.
Here we show
that in a {\it particle description} of an FKPP system 
quenching can occur.

Still considering premixed particles in a closed vessel, let us introduce
the possibility that a stable particle ($B$) can turn into
an unstable one ($A$). That is, beyond the autocatalytic
reaction~(\ref{eq:autocatalytic}), we introduce 
a new  reaction
\begin{equation}
B \,\, \stackrel{q}{\longrightarrow} \,\,  A  \label{quenching}\,,
\end{equation}
where $q$ is its rate.
Therefore we have the following
reaction rules:
\begin{itemize}
\item if $C_\alpha(t) = 0$ then $C_\alpha(t+\Delta t) = 1$ with
  probability $P_{AB} = W_{AB}  \Delta t$
\item if $C_\alpha(t) = 1$ then $C_\alpha(t+\Delta t) = 0$ with
  probability $Q_{BA} = W_{BA}  \Delta t$
\end{itemize}
$\WAB$ is the same of the previous section, while
$W_{BA} = q$ does not depend on the interaction radius $R$,
since it is a single particle property.

\begin{figure}
\epsfig{figure=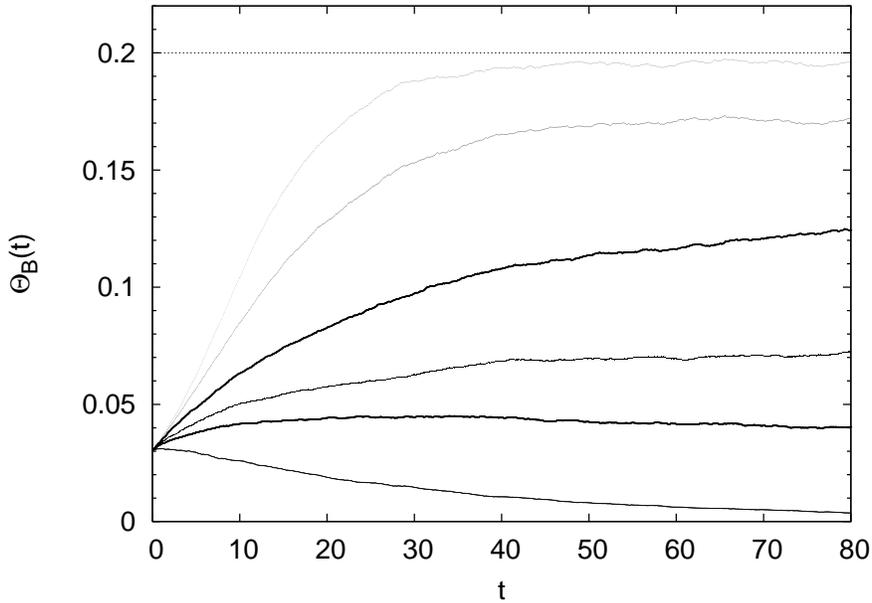,clip=true,width=12truecm,angle=0}
\caption{Time evolution of $\Theta_B(t)$ for $R=0.5,0.1,0.05,0.03,0.02,0.005$ 
(from top to bottom), $N=1000$, $D=0.001$; the straight line is the continuum limit asymptotic value $1-q/p$, 
with $p=1 \, , \, q=0.8$. For $R < 0.02$ the reaction quenches.}
\label{fig:thetaB_t}
\end{figure}

\begin{figure}
%\begin{center}
\epsfig{figure=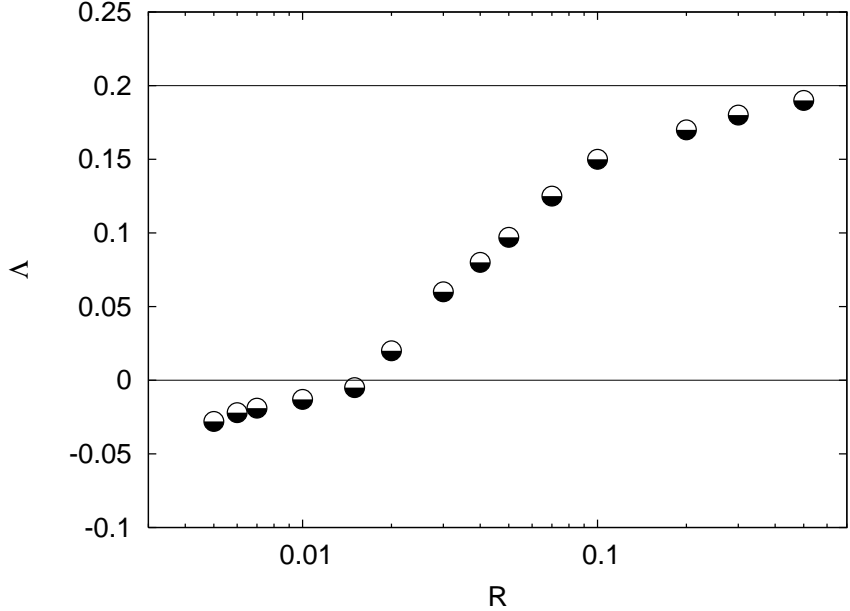,clip=true,width=12truecm,angle=0}
\caption{Inverse characteristic time of the reacting process
$\Lambda=p_R(N)-q$ as a function of the interaction radius $R$ 
(with $p=1 \, , \, q=0.8$); $N=1000$, $D=0.001$. For large values of $R$, $\Lambda$ tends
to the continuum limit value $p-q$; for $R < 0.02$ $\Lambda$ becomes
negative, highlighting the emergence of the quenching phenomenon.}
%\end{center}
\label{fig:Lambda_R}
\end{figure}

The renormalized description  of this model is  given by
\begin{equation}
\frac{d \Theta_{B} (t)}{dt}=p_R(N) \Theta_{B} (1 -\Theta_{B})- q \Theta_{B},
\label{renormalizeddecay}
\end{equation}
whose  solution is  
\begin{equation}
\Theta_B (t) = \Theta_{AS}\frac{\Theta_B (0) e^{\Lambda t}}
      {\Theta_{AS} + \Theta_B (0) (e^{\Lambda t} -1 )}.
\label{sol2}
\end{equation}
with $\Lambda= p_R(N) - q$ and $\Theta_{AS}=1 - q/p_R(N)$.
Two different scenarios now appear. 
If $p_R (N) < q$ for all $N$ the reaction finishes.
On the other side,
when $p_R(N) > q$ we have a similar behavior as in the case
with $q=0$.
In fig. \ref{fig:thetaB_t} we show $\Theta_B$ vs $t$ for
different values of $R$. It is apparent that 
for large $R$ the system behaves similarly to the case $q=0$ (including
the continuum limit for the long time  value of the concentration,
$1-p/q$, for large $R$).
However, for $R$ small enough the concentration asymptotically vanishes, 
that is we have a quenching phenomenon. 
In Fig.~\ref{fig:Lambda_R} where we plot $\Lambda$ vs $R$ (obtained analogously to
Fig.~\ref{fig:versusR}, i.e., fitting the analytical solution
to the numerical results).
For large $R$ we approach the continuum limit
and $\Lambda \to p - q$,
while for small $R$ we have quenching corresponding
to negative values of $\Lambda$.

This is a relevant result, 
entirely due to the role of the interaction
radius, $R$, which reflects
the discrete character of the model 
in the quenching mechanism.
Let us note that, at variance with the results in the
previous section (which are just quantitative
changes with respect to the continuous equation),
now the discrete nature of the system is able to produce
a feature (the quenching) which is absent in the continuum
limit~\cite{Roquejoffre}.

\section{Front properties}
\label{sec:front}

In the previous sections we have studied the dynamics of 
interacting particle systems in a closed container. 
We now focus on a different configuration, corresponding 
to well-separated chemicals in an open domain,  and we  investigate 
evolution properties, such as front speed and thickness \cite{sokolov,LN98}, in
terms of the interaction radius.
%The interaction radius have revealed to take a very relevant role,
%showing a quite strong influence on the reaction rate, and
%determining whether
%a FKPP continuum limit may occur and can produce a quenching
%of the dynamics (when  the reaction $B \to A$ is permitted).

In this section we take $L_y=1$,  $L_x =5$, with 
periodic boundary conditions in the $y$ direction, and
rigid walls in the $x$ direction. The burnt (type $B$)
particles are initially concentrated in the leftmost part of the
system, so that a propagating reaction front, from left to right,
develops. The reaction term we use is
just the autocatalytic one~(\ref{eq:autocatalytic}), i.e.,
$q=0$. We separately study the front speed and
the front thickness.

\subsection{Front speed}
\begin{center}
\begin{figure}
\epsfig{figure=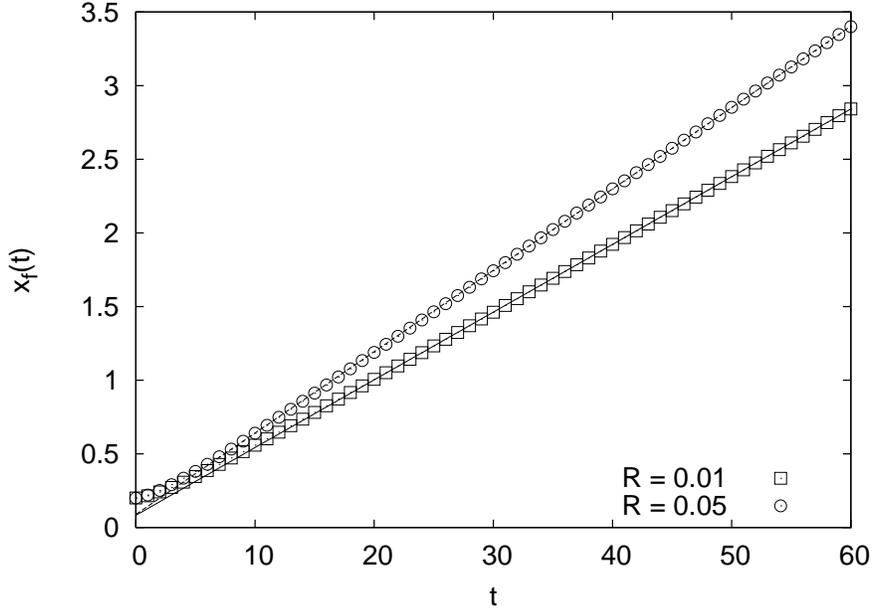,clip=true,width=12truecm,angle=0}
\caption{$\xf$ vs $t$. The system parameters
are: $L_x=5$, $L_y=1$, $D=0.001$, $N=5000$, and the initial number of
$B$ particles is $200$.
The slopes of the straight
lines represent the  front speed.}
\label{fig:evolution}
\end{figure}
\end{center}

We can  define the instant front position as
\begin{equation}
\xf(t) =L_x \frac{N_B (t)}{N}\,,
\label{eq:frontposition}
\end{equation}
and the front speed: 
\begin{equation}
\vf \simeq \frac{\xf(t)}{t}
\label{eq:frontspeed}
\end{equation}
which is computed after a transient and before the complete
saturation is approached. In Fig.~\ref{fig:evolution} we show
$\xf (t)$ vs time for different values of $R$. The $\vf$ is
obtained as the slope of the best fit to the curves in the 
proper interval.  

\begin{center}
\begin{figure}
\epsfig{figure=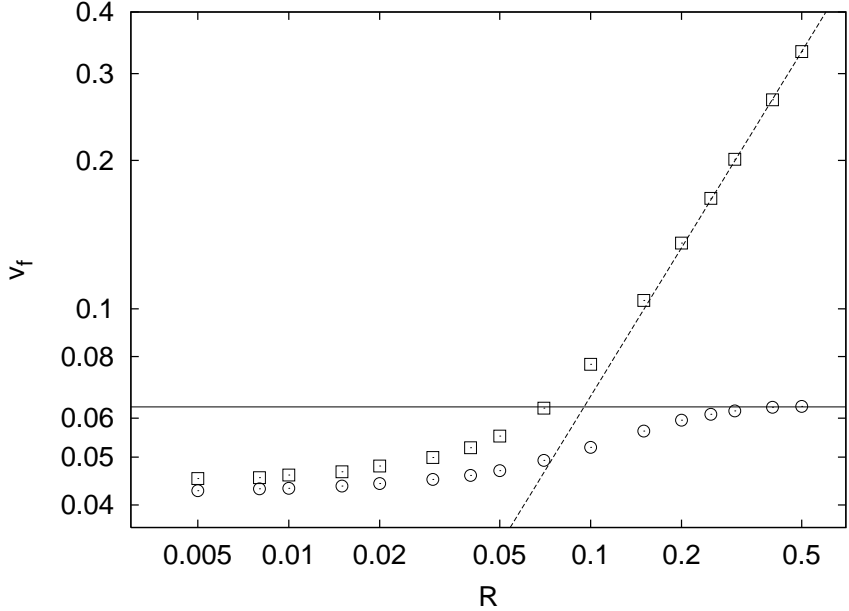,clip=true,width=12truecm,angle=0}
\caption{($\boxempty$) Front speed $\vf$ versus $R$
and ($\circ$) the prediction of formula~(\ref{eq:fkppfrontspeed}),
and $p_R (N)$ computed in closed domains with
initially separated particles distribution. 
The horizontal line is the value in the FKPP
continuum limit, while the dashed line is the behaviour $v_f \simeq R$.
The parameters are the same as Fig.~\ref{fig:evolution}.
The Kuramoto length is $l_k=0.045$.}
\label{fig:frontspeed}
\end{figure}
\end{center}

We expect that, via the
renormalized description 
of the FKPP equation, that is
Eq. ~(\ref{fkpp}) with 
$p$ replaced by $p_R(N)$,
the front speed of the particle model at
varying $R$ should be
\begin{equation}
   v_0 = 2\sqrt{Dp_R(N)}\,.
   \label{eq:fkppfrontspeed}
\end{equation}
We saw that in closed basins  
different initial conditions on the distribution
of particles select different
$p_R(N)$'s, see Figure~\ref{fig:compinitcond}.  Therefore, for the
study of front propagation the proper $p_R(N)$ is that one computed in
the case of initially separated particles 
distribution (symbol $(\boxempty)$
in Figure~\ref{fig:compinitcond}).
The numerical results, reported in Fig.~\ref{fig:frontspeed},
confirm our prediction at least for small $R$, i.e., 
the front speed behaves as in the FKPP case 
(Eq.~(\ref{eq:fkppfrontspeed})).

\begin{center}
\begin{figure}
\epsfig{figure=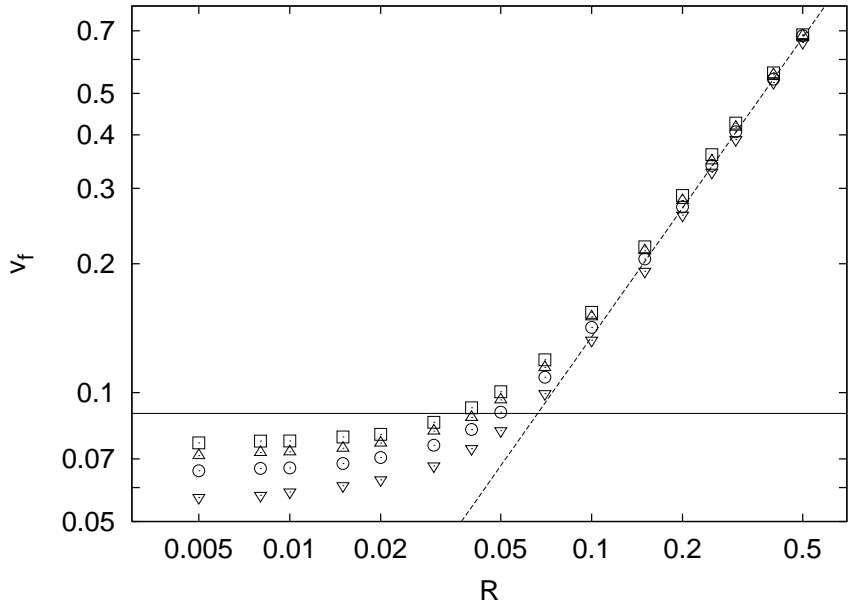,clip=true,width=12truecm,angle=0}
\caption{The measured front speed $\vf$ versus $R$ for ($\triangledown$) 
$N=5000$, ($\circ$) $N=10000$, ($\boxempty$) $N=20000$ 
and ($\triangle$) $N=40000$. The horizontal line is the value in the FKPP
continuum limit, while the dashed line is the behaviour 
$v_f \simeq R$.  The Kuramoto length is $l_k=0.045$.}
\label{fig:confrontspeedN}
\end{figure}
\end{center}

However, the large discrepancy observed for large $R$
cannot be explained by a simple difference
in the initial particle distribution.
This difference arises because the
interaction radius becomes 
larger than the Kuramoto length
\begin{equation}
  l_K =\sqrt{\frac{2 D}{p}}\,,
  \label{eq:kuramoto}
\end{equation}
and therefore the continuum FKPP limit does not hold. 
Indeed, in particle systems, when $R \le l_k$
the interaction term establishes a connection 
between regions containing  $A$ particles and regions containing
$B$ particles that in the classical FKPP equation
could not be connected.
Therefore, when $R \geq l_K$, it is not possible to obtain
the continuum FKPP limit~(\ref{fkpp}) 
even with an arbitrarily large number of particles.
In Figure~\ref{fig:confrontspeedN} it is shown the
front speed at varying $R$ for various $N$.
One can observe that 
at increasing $N$ for small $R$ the front speed
approaches the FKPP value, while for large
$R$ the front speed does not depend on $N$
and the value is definitely different from
the FKPP value.

\begin{center}
\begin{figure}[h!]
\epsfig{figure=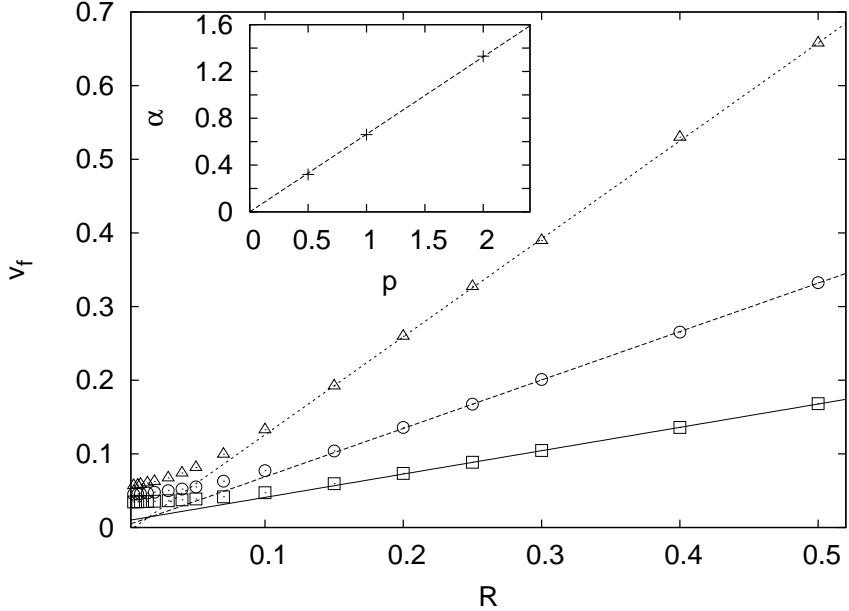,clip=true,width=12truecm,angle=0}
\caption{Large $R$ behaviour of $\vf$ in the cases: 
$p=0.5$ ($\boxempty$), $p=1.0$ ($\circ$) and $p=2.0$ ($\triangle$).
In the inset the slope of the linear fit, $\alpha$ (see 
Eq.~(\ref{eq:alfabehaviour})), is shown as a function of $p$.}
\label{fig:fsvaryp}
\end{figure}
\end{center}

A simple argument explains the behavior of $\vf$
for large $R$.
The front speed 
is proportional to the front length
times the reaction rate, e.g.,  in  the FKPP equation
$v_0 = 2\sqrt{Dp} \propto p  \sqrt{D/p} \propto p l_k$\,.
When the interaction radius is greater than the Kuramoto length
it is reasonable to expect that the front length becomes proportional 
to $R$ and so the front speed: 
\begin{equation}
  \vf \propto p_R(N)  R = \alpha R\,\qquad\mbox{when}\qquad
  R \gg l_k\,,
   \label{eq:vfRgrandi}
\end{equation}
in agreement with the results shown in
Figures~\ref{fig:frontspeed},~\ref{fig:confrontspeedN} 
and in Figure~\ref{fig:fsvaryp} for various $p$. 
In particular, in the inset of Figure~\ref{fig:fsvaryp}
one can see the behaviour of $\alpha$ as a function of $p$:
\begin{equation}
   \alpha(p) = a  p\,,
   \label{eq:alfabehaviour}
\end{equation}
where $a$ is a constant.
This is not surprising because $p$ is the continuum limit
for the reaction rate which is reached asymptotically 
by the particle system, i.e., $p_R(N) \to p$ for large $R$.

\subsection{Front thickness}
\label{sec:thickness}

\begin{center}
\begin{figure}
\epsfig{figure=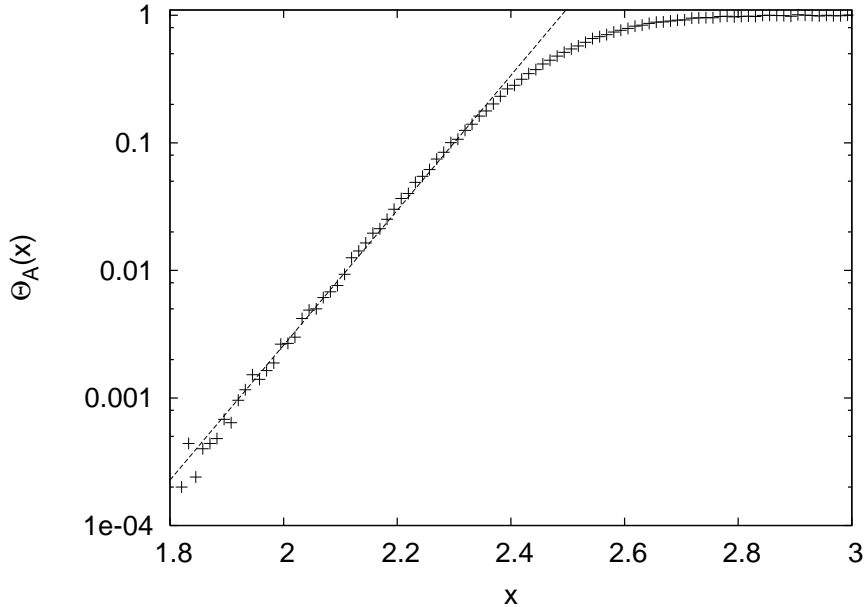,clip=true,width=12truecm,angle=0}
\caption{The shape of the front $\tilde\Theta_A(x)$ $(+)$ and the exponential
fit of equation~(\ref{frontshape}) (dashed line).}
\label{fig:frontthickfit}
\end{figure}
\end{center}

As a further confirmation of previous results,
we investigate the behaviour of front thickness at varying $R$.
Note that in the continuum limit there are many ways to compute the front
thickness of a propagating front~\cite{ciccio}. In the particle case, however,
it is not obvious how to define a front profile. We proceed by 
defining an averaged field that resembles the front shape.
Essentially this is a histogram over particle positions.
Fixing our attention on $A$ particles, we define
\begin{equation}
        \tilde\Theta_A(x,\Delta x, t) = \frac{N_{x,\Delta x}(A,t)}
                                             {N\Delta x}
        \label{fronte}
\end{equation}
where $N_{x,\Delta x}(A,t)$ counts the number of $A$ particles whose
$x$ coordinate lays between $x$ and $x+\Delta x$. When the number of
particles is large the value of $\Delta x$ could be taken arbitrarily
small whereas, in general, $\Delta x$ has to be small, but at the same
time large enough in order to avoid large fluctuations in $N_{x,\Delta
x}(A,t)$.  We use a relatively small $\Delta x$ (few $d_m$) and we
average $N_{x,\Delta x}(A,t)$ over many different realizations.
The front shape of an FKPP system behaves as
\begin{equation}
        \tilde\Theta_A(x,\Delta x; t) \sim \exp((x-\xf(t))/l_A),
        \label{frontshape}
\end{equation}
where $l_A$ is the front thickness, and $\xf(t)$ the front position
at time $t$.
In Figure~\ref{fig:frontthickfit} it is shown the 
exponential behaviour of the front profile and
the fit obtained from of Eq.~(\ref{frontshape}). 
In particular Eq.~(\ref{frontshape}) works well 
in the central region of the front, 
i.e., where corrections due to the particle nature
of the system are less important. 

%Another way to compute the front length goes through
%measuring  the region in which the field
%$\tilde\Theta_A(x,\Delta x; t)$ has values between 
%$0.1$ and $0.9$, $\Gamma_A$.
%This is a reasonable measure of the region in which
%the reaction happens. We have realised that
%both measurments give rise to similar results. 

\begin{center}
\begin{figure}
\epsfig{figure=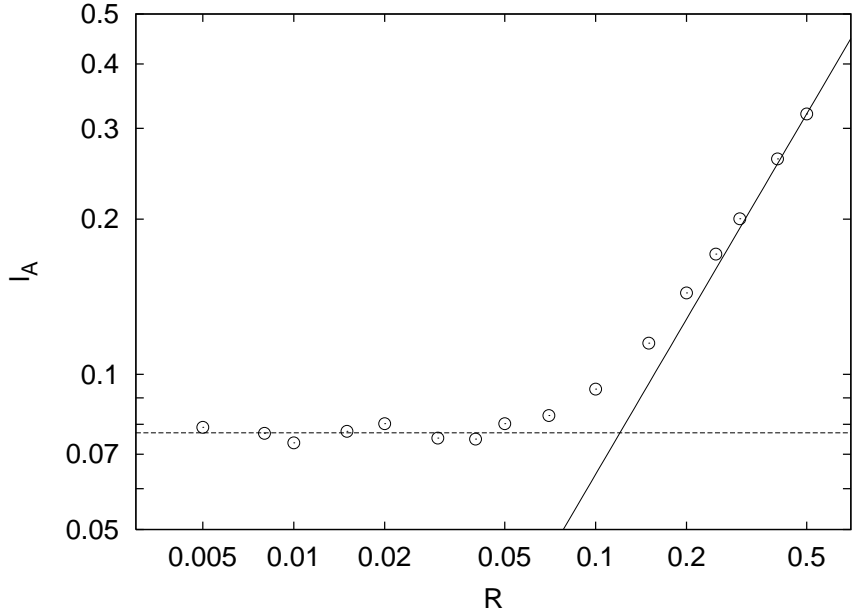,clip=true,width=12truecm,angle=0}
\caption{
Front thickness $l_A$ versus $R$ measured in the particle model ($\circ$).
The dashed line corresponds to a  constant value of the front speed
(see fig.\ref{fig:confrontspeedN}),
while the full line is the behaviour $l_A \simeq R$.
The value of the Kuramoto length scale 
is $l_K=0.045$. }
\label{fig:frontthick}
\end{figure}
\end{center}

Other measurements of the front profile provide
similar results.
In   Figure~\ref{fig:frontthick} we plot the front
thickness, $l_A$, computed for different values of $R$.
Again, for  $R$ smaller than the Kuramoto length
the front thickness is constant, while for values of $R$
greater than $l_k$ the front thickness behaves
as $l_A \propto R$. This result confirms the assumption
of equation~(\ref{eq:vfRgrandi}). The constant value reported 
in Figure~\ref{fig:frontthick} (the dashed line)
is only an indicative value to show that for $R < l_k$ 
the front thickness is constant, and it is not the FKPP value 
of the front width.

\section{Summary and conclusions}
\label{sec:conclusion}

In this work we studied the effects of the discrete-particle character
in an autocatalytic reacting system, described in terms of 
chemical dynamics where two types of Brownian particles 
interact when they are at a distance smaller than a certain radius $R$. 
We have shown that in a suitable continuum limit
the system is equivalent to  an FKPP model for
the concentration of particles, and we have focused
on the differences that arise when the conditions
for this limit are not fullfilled. 

The continuum limit holds if
some relations among the relevant length
scales
(the interaction radius, the mean distance between particles
and the Kuramoto length)
 of the problem are satisfied.
For  well-premixed initial conditions 
of the particles distribution
only the first
two lengths play a role. However, for front propagation, 
 which requires a separated initial
distribution of particles, also the Kuramoto
length  is important.

We have also considered the modified chemical dynamics
$ A + B \,\, \stackrel{ p }{\longrightarrow} \,\, 2B$,
and
$B \,\, \stackrel{q}{\longrightarrow} \,\,  A$, 
where, at variance with the continuum limit,
one can have the possibility of quenching, for small
values of $R$. This is due to the particle nature
of the model, which induces a qualitative difference
with respect to the continuous description. 

 Moreover, in the context of front propagation, 
we have shown that particular conditions exist such that, 
increasing the particle density, the system reaches
a continuum limit which is definitely different
from the continuum FKPP limit.

We conclude noting  that many biological systems are
characterised by the two main ingredients of our work: the 
minimal distance for the interaction,
and the exiguity of the number of organisms~\cite{flierl,bioref}.
We hope
that our work helps to clarify some shortcomings arising when a
macroscopic description is attempted.

%\appendix
%\section{Numerical stuff}
%The key point in the simulation of moving particle
%that interacts is the chose of $\Delta t$.
%There are two inequalities that must be satisfied
%\begin{itemize}
%\item[1)] the diffusive shifting in a $\Delta t$ has to be
%smaller than $R$: $\sqrt{D\Delta t} \ll R$
%\item[2)] the probability to change type of particle
%has to be smaller than $1$: $\WAB  \Delta t \ll 1$
%\end{itemize}

\section{Acknowledgments}

We have benefited from a MEC-MIUR joint program (Italy-Spain Integral Actions). 
C.L. acknowledges  support 
from FEDER and MEC (Spain),
through project CONOCE2 (FIS2004-00953). 
A.V. and D.V. acknowledge support from PRIN-MIUR 
project "Dinamica Statistica di sistemi a molti
e pochi gradi di libert\`a".
We warmly acknowlege Emilio Hern\'andez-Garc\'\i a 
and Simone Pigolotti
for a critical
reading of the manuscript.

%\end{twocolumns}


\begin{thebibliography}{99}

\bibitem{murray}
J.D. Murray, {\it Mathematical Biology}, Springer-Verlag,
Berlin (1993).

\bibitem{flierl}
G. Flierl, D. Grunbaum, S. Levin, and D. Olson,
J. Theor. Biol. {\bf 196}, 397 (1999).

\bibitem{tel}
T. T\'el, A. de Moura, C. Grebogi and G. K\'arolyi,
Phys. Rep., {\bf 413}, 91 (2005).

\bibitem{kolmogorov} 
A.N.~Kolmogorov, I.~Petrovskii and N.~Piskunov, 
Bull. Univ. Moscow, Ser. Int. A, {\bf 1}, 1 (1937).

\bibitem{fisher} 
R.A.~Fisher, 
Ann. Eugenics, {\bf 7}, 353 (1937).

\bibitem{saarloos}
W. van Saarloos,
Phys. Rep. {\bf 386}, 29 (2003).

\bibitem{xin}
J. Xin, SIAM Review {\bf 42}, 161 (2000).

\bibitem{aronson} 
D.G.~Aronson and  H.F.~Weinberger, 
Adv. Math., {\bf 30}, 33 (1978).

\bibitem{young}
W.R. Young, A.J. Roberts, G. Stuhne, Nature {\bf
412}, 328 (2001).


\bibitem{lopez}
 E. Hern\'andez-Garc\'\i a and C. L\'opez, Phys. Rev. E {\bf
70} 016216 (2004);\\
C. L\'opez and E. Hern\'andez-Garc\'\i a, Physica D {\bf
199}, 223 (2004).



\bibitem{discrete1}
C. R. Doering, C. Mueller, and P. Smereka,
Physica A {\bf 325}, 243 (2003).


\bibitem{brunet}
E. Brunet and B. Derrida, Phys. Rev. E {\bf 56}, 2597 (1997).

\bibitem{panja}
D. Panja,
Phys. Rep. {\bf 393},  87 (2004).

\bibitem{kaneko}
Y. Togashi and K. Kaneko, 
Phys. Rev. Lett. {\bf 86}, 2459 (2001);\\
Y. Togashi and K. Kaneko,
Phys. Rev. E {\bf 70}, 020901 (2004).


\bibitem{plankton}
A.W. Visser and U.H. Thygesen,
J. Plankton Res. {\bf 25}, 1157 (2003);
C. L\'opez, 
Phys. Rev. E {\bf 72}, 061109 (2005).

\bibitem{Constantin}
N. Vladimirova, P. Constantin, A. Kiselev, O. Ruchayskiy
and L. Ryzhik,
Combust. Theory Modelling {\bf 7}, 487 (2003).


\bibitem{Roquejoffre}
J.M. Roquejoffre,
Arch. Rat.Mech. Anal. {\bf 117}, 119 (1992);\\
J.M. Roquejoffre,
Ann. Inst. H. Poincar\'e, Anal. Nonlin. {\bf 14}, 499 (1997).

\bibitem{sokolov}
J. Mai, I.M. Sokolov, and A. Blumen,
Europhys. Lett {\bf 44}, 7 (1998);\\
J. Mai, I.M. Sokolov, and A. Blumen,
Phys. Rev. E {\bf 62}, 141 (2000).

\bibitem{LN98}
A. Lemarchand and B. Nowakovski, 
Europhys. Lett. {\bf 41}, 445 (1998).

\bibitem{ciccio}
 M. Abel, A. Celani, D. Vergni and A. Vulpiani,
Phys. Rev. E {\bf 64}, 046307 (2001).

\bibitem{bioref}
R. Durrett and S.A. Levin, 
 Theoretical Population Biology {\bf 46},  363 (1994).

\end{thebibliography}
\end{document}